\documentclass[twocolumn,showpacs,preprintnumbers,amsmath,amssymb,superscriptaddress]{revtex4}
\input epsf
\epsfverbosetrue
\def\be{\begin{equation}}
\def\ee{\end{equation}}
\def\bea{\begin{eqnarray}}
\def\eea{\end{eqnarray}}

\newcommand{\lsb}{\left[}
\newcommand{\rsb}{\right]}
\def\vecx{\underline{x}}

\def\veck{\underline{k}}
\def\kdotx{\underline{k}\cdot\underline{x}}
\def\momvol{\frac{d^{3}\underline{k}}{(2\pi)^{3}}}
\def\nn{\nonumber}
\newcommand{\lrb}{\left(}
\newcommand{\rrb}{\right)}
\newcommand{\ba}{\begin{array}}
\newcommand{\ea}{\end{array}}

\def\adag-k{a^{\dag}_{-k}}
\def\a-k{a_{-k}}

\def\ddag-k{d^{\dag}_{-k}}
\def\d-k{d_{-k}}

\def\Bdag-k{A^{\dag}_{\theta_k}}
\def\B-k{A_{\theta_{-k}}}

\def\C-k{D_{\theta_{-k}}}
\def\Cdag-k{D^{\dag}_{\theta_k}}

\def\F-k{F(k)}
\def\Fdag-k{F^{\dag}(-k)}
\def\bmat{\lrb\begin{array}{c}}
\def\emat{\end{array}\rrb}

\begin{document}
\title{Isospin Squeezed States , Disoriented Chiral condensates and Pion Production: A Dynamic Group Theoretical Approach}
\author{B.Bambah}
\affiliation{ School of Physics \\ University of
  Hyderabad, Hyderabad-500 046,India }
\author{C.Mukku}
\affiliation{Department of Mathematics \\and\\ Department of Computer Science \& Applications\\Panjab University, Chandigarh-160014,India}
\begin{abstract}
We make a complete dynamical study of Isotopic spin
conservation effects on the multiplicity distributions of both
hard and soft  pions emitted in a quark gluon plasma undergoing a
non-equilibrium phase transition.
\end{abstract}
\maketitle \large
\section{Introduction}
Quantum optical analogies have long been useful tools in
describing multipion production in hadronic collisions
\cite{knox,giov,carruthers,bambah,weiner}. Quantities such as pion
multiplicity distributions and correlations among identical and
charged pions have been successfully explained on a
phenomenological level by exploiting the similarity between light
and bosonic particles such as pions. This has been very successful
for systematizing data in high energy collisions since the
"pre-QCD" days ,when the concept of coherent pion production was
introduced \cite{horn, botke,Waka} . Since then,  many
distributions which have their origin in photon-counting
experiments have found an application in particle physics
\cite{weiner}. In particular the negative binomial distribution
\cite{giov}, the Perina-McGill distribution \cite{biya}, and the
squeezed coherent distribution \cite{bambah,Amado} have proved very
useful in various high energy applications. Recently these quantum
optical states have been revived as possible states for the
disoriented chiral condensate, which is conjectured to occur in
relativistic heavy ion collisions \cite{kowal,bj1,rw}.

The distinguishing factors between optical and pionic physics are
the conservation laws and final state interactions between pions
\cite{Mart}. Thus,  in translating the quantum optical analogy to
pion production,  care has to be taken to preserve the
conservation laws such as isospin, charge etc.  obeyed by the
underlying interactions \cite{andreev,botke2,skagerstam}.
 This is done by imposing constraints which exploit the group theoretic
structure of the symmetries and involve projection operator techniques
 \cite{hwa}.
 In particular for strong interactions where isospin is conserved, coherent states
 of pions are obtained by globally by projecting out a particular
isospin state from a general number operator coherent state. Such
a global projection assumes that all pions are assigned the same
momentum function and thus  isospin dependence of event by event
quantities like momentum dependence of pion correlations (HBT
effect) and back-to-back particle-antiparticle correlations (PAC)
cannot be described in this approach. One has to do a full quantum
field theoretical treatment in momentum space and find momentum
dependent isospin variables to give a full event by event
description of the isospin coherent state.

 Recently a number of papers in   literature have addressed the
question of isospin conservation in the context of pion production
associated with the disoriented chiral condensate \cite{hwa,Hiro,suzuki,sigal}. These involve the construction of the isospin
squeezed states. However, these papers do not  make any proposal
for the dynamical origin for the anomalous production of particles
in relation to the squeezed states. This has motivated us  to look
deeper into the question of whether isospin squeezed states  can
be dynamically generated in models for pion production in certain
high energy processes. Furthermore,  only the isospin structure at
zero relative momentum  of pions has been investigated \cite{Hiro}. Our
approach is the most general one which has the results of these
papers as special cases, and is more suitable for the dynamical
origin of pion distributions seen in physical processes. It can
also be easily generalized to include disoriented chiral
condensates., so that isospin invariance in DCC production does
not have to be compromised.
\section{The Pion Hamiltonian} The foundations for the model used
are laid in our previous paper \cite{bm1} , hitherto referred to
as I.  In the next two sections , to make this paper complete and also to relieve the reader from the ardrous task of
continually referring to I, we outline the salient features and results derived in I.
Since this is the basis for our subsequent discussion and results, the repetition is not redundant but useful.

In I,  we used a background field analysis of the
$O(4)$ sigma
 model  with symmetry breaking keeping one-loop quantum corrections to construct the most
 general Hamiltonian for the evolution , formation, decay  of the DCC  in an external,
expanding metric.

 The complete action for the  pion-sigma part of the  O(4) sigma model  which preserves chiral symmetry is
given by
\be
S = \int d^4 x{1\over2}\, ( \partial_\mu {\vec \pi}\, \partial^\mu
{\vec \pi} +
\partial_\mu \,\Sigma \partial^\mu \Sigma) -
{\lambda\over 4}\, ( {\vec \pi}^2 + \Sigma^2 -f_{\pi}^2)^2 \ee

Where, the chiral field is
\begin{equation}
\Phi(r,t)=\Sigma(r,t)+ i\bf{\vec{\tau}}\cdot\bf{\vec{\pi}}(r,t).
\end{equation}
The true vacuum is $<\Phi>^2=<\Sigma>^2+<\pi>^2=f_{\pi}^2$ and is
taken traditionally as $<\Sigma>=f_{\pi}$and $<\pi>=0$. 
The choice of this
vacuum implies that chiral symmetry is spontaneously broken. 
Note
that the fermionic part of the original sigma model Lagrangian has
been neglected here. This is allowed since our focus is on the
condensate formed in the symmetry broken phase where quark degrees
of freedom are already confined. The explicit breaking of symmetry
is implemented in the Lagrangian to give the pions mass through
the presence of a term linear in the sigma field.The action
becomes

\be
 S=\int d^{4}x
\left[{1\over2}(\partial_\mu\phi^a\partial_{\mu}\phi_{a}
-{\lambda\over4}(\phi^a\phi_a-v^2)^2+\epsilon\Sigma \right] .
\ee
where  $f_{\pi}$ is replaced by a general parameter $v$ , which in the
limit $\epsilon\longrightarrow0$ goes to $f_{\pi}$. 
Requiring
the minimum to be still $(0,f_{\pi})$ ,  to leading order we have 
\be
v^2=f_{\pi}^2-\frac{m_{\pi}^2}{\lambda} \ee and the pion and sigma  masses  are
$m_{\pi}^2=\frac{\epsilon}{f_{\pi}}$ and $m_{\sigma^2}=2\lambda
f_{\pi}^2+m_{\pi}^2$.\\
  $f_\pi=92MeV$ is the pion decay
constant and the meson masses are
 $m_\pi=138MeV/c^2$ and $m_\sigma=600MeV/c^2$.

Although , to give the pions mass,  we have explicitly broken the
symmetry, if the explicit symmetry breaking is very small compared
to the relevant scale of QCD then it is still a good approximation
to apply the notion of spontaneously broken symmetry.  the symmetry breaking term tilts the double well potential given 
in (3). As long as
the potential is tilted only slightly the pions are much softer
than the sigma so the effect due to the spontaneous symmetry
breakdown of chiral symmetry dominates the dynamics if $\epsilon$ is
small.

The formation  of the DCC takes place when,  in a rapidly cooling
and expanding plasma, the expansion rate of plasma is greater than
the rate at which the field evolves from a state of restored
symmetry to an equilibrium state of broken symmetry \cite{bj1,rw}. This
evolution of the field can be implemented by making  the expectation value of $\Phi$ a
function of time. Such a study has been carried out in reference
{\cite{bm1} in great detail. An additional feature in the study of
the formation of the DCC is that of the expansion of the plasma --
two scenarios exist, the first being that of a sudden quench,
where $<\Phi>$ changes from $0$ to $f_{\pi}$ instantaneously and
the second in which the system goes through a meta-stable,
disordered vacuum typically represented by:
\be
\Sigma=f_{\pi}Cos(\theta);
\;\;\vec{\pi}=f_{\pi}\vec{n}Sin(\theta). \ee The system then
relaxes by quantum fluctuations to an equilibrium configuration.
Here $\theta$ measures the degree of disorientation of the
condensate. 

Some of the traditional signals of the DCC are the enhancement of
 low momentum pion modes \cite{rw} and the anomalous charged
 to pion ratio of the soft pions \cite{mart}.
 The study of the DCC requires the further analysis of
the pions produced by the decaying plasma and therefore it is
imperative that the isospin structure , which is such an integral
part of pion studies,  has also to be incorporated. Studies carried
out thus far have concentrated on classical isospin structure. We
look at the isospin structure at the quantum level through the
construction of an effective mean field Hamiltonian which allows
us to examine the dynamical effects of isospin conservation. This
provides us with a framework for examining the effects of isospin
conservation on the pion multiplicities.

In \cite{bm1} we have studied the effects of an $so(4)$ sigma
model with spontaneous symmetry breaking in a spherically
symmetric and homogeneously expanding plasma. The line element is
the FRW metric:
\be
ds^{2}=dt^{2}-a(t)^{2} d\vec{x}^{2}, \ee where $a(t)$ is the
expansion parameter . We treated the quantum field as a
fluctuation around the  general classical background field
 parameterized by three angles:
\begin{widetext}\be<\Phi>=\lrb \ba{c}
f_{\pi}Cos(\rho)Sin(\theta)Sin(\alpha)\\f_{\pi}(\rho)Sin(\theta)Cos(\alpha)\\f_{\pi}Sin(\rho)Sin(\theta)\\f_{\pi}Cos(\theta)\ea
\rrb=\lrb \ba{c}
v_+\\v_-\\v_3\\\sigma \ea \rrb \ee \end{widetext}
where $v_+,v_-$ are the vacuum expectation values of the charged pions and $v_3$  of the neutral pion .
Without DCC formation these would be zero.
In order to consider all the special cases that are possible in a
transparent way, we simplified  the parameterizations of the
possible form for the background field to two angles, $\theta$ and
$\rho$ by letting $\alpha=\frac{\pi}{4}$.
Then,
$v\pm=\frac{f_{\pi}}{\sqrt{2}}Cos(\rho)Sin(\theta)$,\hspace{0.5cm}$v_{3}=f_{\pi}
Sin(\rho)Sin(\theta)$,\hspace{0.5cm} and
\hspace{0.5cm}$\sigma=f_{\pi}Cos(\theta)$.

 The quantum
Hamiltonian was derived using  a Fourier mode decomposition of the
fields and momenta through the definitions
 \bea
\pi_{0}(\vecx,t)&=& \int \sqrt{\frac{1}{2\omega_{\pi}}}\momvol
\{a_{\veck}e^{i\kdotx} +a_{\veck}^{\dag}e^{-i\kdotx}\} \nn \\
 \pi_{-}(\vecx,t)&=&\int \sqrt{\frac{1}{2\omega_{\pi}}}\momvol
\{b_{\veck}e^{i\kdotx} +c_{\veck}^{\dag}e^{-i\kdotx}\}  \nn \\
\pi_{+}(\vecx,t)&=&\int \sqrt{\frac{1}{2\omega_{\pi}}}\momvol
\{c_{\veck}e^{i\kdotx} +b_{\veck}^{\dag}e^{-i\kdotx} \} \nn \\
 \Sigma(\vecx,t)&=& \int \sqrt{\frac{1}{2\omega_\Sigma}}\momvol
\{d_{\veck}e^{i\kdotx} +d_{\veck}^{\dag}e^{-i\kdotx}\}\nn \\ \eea where
\bea
\frac{\omega^{2}_{\pi}(k)}{a^{6}}\equiv\frac{\omega^{2}_{\pi_0}(k)}{a^{6}}=\frac{\omega^{2}_{\pi_\pm}(k)}{a^{6}}=(m_{\pi}^{2}+\frac{\veck^{2}}{a^{2}})\\
\nn
\frac{\omega^{2}_{\Sigma}(k)}{a^{6}}=(m_{\Sigma}^{2}+\frac{\veck^{2}}{a^{2}})
\eea and  we also define \bea
\frac{\Omega^2_{\pi_0}-\omega^2_{\pi_0}}{a^6}&=&\lambda[(<\Phi>^2-v^2)+2v_3^2]
\nn \\ \frac{\Omega^2_{\pi
{\pm}}-\omega^2_{\pi_{\pm}}}{a^6}&=&\lambda[(<\Phi>^2-v^2)+2v_+v_-]
\nn
\\
\frac{\Omega^2_{\Sigma}-\omega^2_{\Sigma}}{a^6}&=&\lambda[(<\Phi>^2-v^2)+2\sigma^2]
\eea and
\be
2v_+v_-+v_3^2+\sigma^2=f_{\pi}^2. \ee   For the rest of the
section, for brevity, we drop the k,t dependence of the $\Omega$'s
and $\omega$'s. The Hamiltonian that describes the quantum
evolution of a DCC where the corresponding classical condensate
can be in any direction in isospin space.
 is:
\be
H=H_{neutral}+H_{charged}+H_{mixed} \ee where,
\begin{widetext}
\bea
 H_{neutral}&=&\int \momvol
\frac{1}{2}\{
\frac{\omega_{\pi}}{a^3}(a_k^{\dag}a_k+a_ka^{\dag}_k) \nn \\
&+&\frac{\omega_{\pi}}{2a^3}(\frac{\Omega_{\pi}^2}{\omega_{\pi}^2}-1)(a_k^{\dag}a_k+a_ka_k^{\dag}+a_{-k}a_k+a_{-k}^{\dag}a_k^{\dag})
 +\frac{\omega_{\Sigma}}{a^3}(
 d_k^{\dag}d_k+d_kd^{\dag}_k) \nn \\
 &+&\frac{\omega_{\Sigma}}{2a^3}(\frac{\Omega_{\Sigma}^2}{\omega_{\Sigma}^2}-1)(d_k^{\dag}d_k+d_kd_k^{\dag}+d_{-k}d_k+d_{-k}^{\dag}d_k^{\dag})\}
\eea \end{widetext} \begin{widetext}\be
H_{charged}=\int \momvol
\{ \frac{\omega_{\pi}}{a^3}(b_k^{\dag}b_k+c_kc^{\dag}_k)  
+\frac{\omega_{\pi}}{2a^3}(\frac{\Omega_{\pi_\pm}^2}{\omega_{\pi}^2}-1)(b_k^{\dag}b_k+c_kc_k^{\dag}+b_{-k}c_k+c_{-k}^{\dag}b_k^{\dag})\}
\ee\end{widetext}
\begin{widetext} \bea
 H_{mixed}=\int \momvol
\{ \frac{\lambda
a^3f_{\pi}^2cos^2(\rho)sin^2(\theta)}{4\omega_{\pi}}(
 b_k b_{-k}+b_kc^{\dag}_k
 \nn \\
 +c^{\dag}_k b_k+c_k c_{-k}+c^{\dag}_k
 c^{\dag}_{-k}+c_kb^{\dag}_k+b^{\dag}_kc_k+b^{\dag}_kb^{\dag}_{-k})
 \nn \\
+ \frac{\lambda
a^3f_{\pi}^2cos(\rho)sin(\rho)sin^2(\theta)}{\sqrt{\omega_{\Sigma}\omega_{\pi}}}
\nn\\  \lrb  b_k a_{-k}+b_ka^{\dag}_k+c^{\dag}_k a_k+c_k
a_{-k}+c^{\dag}_k
a^{\dag}_{-k}+c_ka^{\dag}_k+b^{\dag}_ka_k+b^{\dag}_ka^{\dag}_{-k}\rrb
\nn \\ + \frac{\lambda
a^3f_{\pi}^2sin(\rho)sin(\theta)cos(\theta)}{\sqrt{\omega_{\pi}\omega_{\Sigma}}}\lrb
 d_k a_{-k}+d_ka^{\dag}_k+d^{\dag}_k a_k+d^{\dag}_k
 a^{\dag}_{-k}\rrb \nn \\
+ \frac{\lambda
a^3f_{\pi}^2cos(\rho)sin(\theta)cos(\theta)}{\sqrt{\omega_{\pi}\omega_{\Sigma}}}
\nn \\ \lrb
 b_k d_{-k}+b_kd^{\dag}_k+c^{\dag}_k d_k+c_k d_{-k}+c^{\dag}_k d^{\dag}_{-k}+c_kd^{\dag}_k+b^{\dag}_kd_k+b^{\dag}_kd^{\dag}_{-k}\rrb
 \} \nn \\
 \eea \end{widetext}

 In order to show the dynamical origin of squeezed isospin states we consider the simplest case   $\theta=0$.
In subsequent communications the other cases will be considered.
$\theta=0$ implies the symmetry breaking takes place in the
$\Sigma$ direction and the  mixed term Hamiltonian , $H_{mixed}$
vanishes. The general problem of evolution of the quantum state of
the DCC with arbitrary orientation in isospin space will be
considered later \cite{maedan}. The total Hamiltonian, $H$ for this special case
reduces to \begin{widetext} \bea
 H&=&\int \momvol
\{ \frac{\omega_{\pi}}{2a^3}(a_k^{\dag}a_k+a_ka^{\dag}_k)\\ \nn
&+&\frac{\omega_{\pi}}{4a^3}(\frac{\Omega_{\pi}^2}{\omega_{\pi}^2}-1)(a_k^{\dag}a_k+a_ka_k^{\dag}+a_{-k}a_k+a_{-k}^{\dag}a_k^{\dag})
\nn \\
 &+&\frac{\omega_{\Sigma}}{2a^3}(
 d_k^{\dag}d_k+d_kd^{\dag}_k)\\ 
\nn &+&\frac{\omega_{\Sigma}}{4 a^3}(\frac{\Omega_{\Sigma}^2}{\omega_{\Sigma}^2}-1)(d_k^{\dag}d_k+d_kd_k^{\dag}+d_{-k}d_k+d_{-k}^{\dag}d_k^{\dag})\}\\
 \nn
&+&
\{\frac{\omega_{\pi}}{a^3}(b_k^{\dag}b_k+c_kc^{\dag}_k)+\frac{\omega_{\pi}}{2a^3}(\frac{\Omega_{\pi}^2}{\omega_{\pi}^2}-1)(b_k^{\dag}b_k+c_kc_k^{\dag}+b_{-k}c_k+c_{-k}^{\dag}b_k^{\dag})\}
 \eea \end{widetext}
 We notice that H has the form of a decoupled
Hamiltonian . This is easy to understand from the $S0(4)$ parent.
The S0(4) vector has been decomposed into four fields
:$\pi_{\pm}$,$\pi_0$ and $\Sigma$ being respectively the charged
pions, the neutral pions and the sigma fields.  H has the
characteristic quadratic su(1,1) structure that is associated with
Hamiltonians that are diagonalised by a squeezing transformation , which was explicitly done in 1.

The parameters of the squeezing transformation in the pion sector are \begin{widetext}:
\be
\mu=Cosh(r_{\pi})=\sqrt{\frac{1}{2}[(\frac{\Omega_{\pi}}{\omega_{\pi}}+\frac{\omega_{\pi}}{\Omega_\pi})+1]}
\; \;\;\;\;\
\nu=Sinh(r_{\pi})=\sqrt{\frac{1}{2}[(\frac{\Omega_{\pi}}{\omega_{\pi}}+\frac{\omega_{\pi}}{\Omega_{\pi}})-1]}.
\ee

\end{widetext} and in the $\sigma$ sector are:
\begin{widetext}
\be
\rho=Cosh(r_{\Sigma})=\sqrt{\frac{1}{2}[(\frac{\Omega_{\Sigma}}{\omega_{\Sigma}}+\frac{\omega_\Sigma}{\Omega_\Sigma})+1]}
\; \;\;\;\;\
\sigma=Sinh(r_{\Sigma})=\sqrt{\frac{1}{2}[(\frac{\Omega_{\Sigma}}{\omega_\Sigma}+\frac{\omega_\Sigma}{\Omega_\Sigma})-1]}.
\ee
\end{widetext}
 $r$ and $r_{\Sigma}$ are the squeezing parameters,
and $\mu^{2}-\nu^{2} = 1$ and $\rho^2-\sigma^2=1$.

Since the sigma field  decouples in this particular Hamiltonian,
it can be analyzed independently of the pion fields. The
diagonalized Hamiltonian can be written as:\begin{widetext} \be H=\int
\momvol\frac{1}{2a^3}\{ \Omega_{\pi}\{(A^{\dag}_{k} A_{k} +
\frac{1}{2}) +( C^{\dag}_{k} C_{k} + B^{\dag}_{k}
B_{k}+1)\}+\Omega_{\Sigma}(D^{\dag}_{k}D_{k}+\frac{1}{2})\}. \ee
where  \end{widetext}\bea A_{k}(t,r)=\mu(r,t) a_{k}+\nu(r,t) a^{\dag}_{-k}\\ \nn
D_{k}(t,r)=\rho(r,t) d_{k}+\sigma(r,t) d^{\dag}_{-k}\\ \nn
C_{k}(t,r)=\mu(r,t) c_{k} +\nu(r,t) b^{\dag}_{-k}\\\nn
B_{k}(t,r)=\mu(r,t) c_{-k} +\nu(r,t) b^{\dag}_{k} \eea

Considering only the pion sector, the diagonalized Hamiltonian $H$
can be converted into a Hamiltonian in terms of quantum fields
corresponding to the operators $A,B,C $ and their adjoints to
obtain a purely quadratic Hamiltonian.
\be
H(t)=\int\frac{d^3k}{(2\pi)^3}\sum_{i=A,B,C}\frac{1}{2}((\frac{\Omega_{\pi}}{a^3})^2\Pi_{i}^2(k,t)+P_{\Pi_{i}}^2(k,t))\ee

The Schroedinger equation for each momentum mode is simply: \be
H_0(k,t)\psi(k,t)=i\frac{d}{dt}\psi(k,t).\ee If we use the
$\Pi$-representation (co-ordinate space representation) for
$\psi(k,t)$, then, the $su(1,1)$ symmetry of the Hamiltonian tells
us that the solution for $\psi(k,t)$ is just a Gaussian. The
equation satisfied by the wave functions for each mode are then
given by: \be
\ddot\psi_{A}(k,t)+\frac{3\dot{a}}{a}\dot{\psi_{A}}+(\frac{\Omega_{\pi}}{a^3})^{2}(k,t)\psi_{A}(k,t)=0.\ee
(similar equations hold for fields B and C) where \be
(\frac{\Omega_\phi}{a^3})^2(k,t)=(\frac{\veck^2}{a^2})+\lambda(<\Phi>^2(t)-
v^2).\ee The expectation values of the number operator for the
neutral pions for each momentum k is given by: \bea
<\psi_{k}(t)|a^{\dag}_{k}a_{k}|\psi_{k}(t)>&=&Sinh^{2}(r) \nn\\
&=&<\psi_{k}|A^{\dag}_{k}(t)A_{k}(t)|\psi_{k}>.\eea
An identical expression holds for the charged scalar fields.

The non-equilibrium transition is carried out as described earlier
by making $<\Phi>$ a function of time,where $<\Phi>$ changes from
$0$ to $f_{\pi}$  either instantaneously (quench) or
adiabatically. In ref \cite{bm1},we have shown how the squeezing
parameter $r_k$
 is related to the competing effects of the expansion rate of the
 plasma and the rolling down time of the system from a state of
 restored symmetry to that of broken symmetry.
 We found that for the quenched limit(fast expansion) the low
 momentum modes are enhanced due to the squeezing parameter being
 very large, whereas for the adiabatic limit (slow expansion) no
 such enhancement occurs and the squeezing parameter is small. This enhancement corresponds to DCC formation.
\section{The su(1,1) Dynamical symmetry and Pion Multiplicity
Distributions}
The su(1,1) symmetry of this Hamiltonian was studied in ref \cite{bm1}. In
terms of $su(1,1)$ generators the Hamiltonian can be written as \begin{widetext}
\begin{eqnarray}
H&=&H_{\pi}+H_{\Sigma}\\ \nn &=&\int \momvol\frac{1}{a^3}
2\Omega_{\pi }((\mu^2+\nu^{2})N+
\mu\nu({{\cal{D}}}+{{\cal{D}}}^{\dag}))+\int
\momvol\frac{1}{a^3}2\Omega_{\Sigma
}((\rho^2+\sigma^{2})N_{\Sigma}+
\sigma\rho({\cal{D}}_{\Sigma}+{\cal{D}}^{\dag}_{\Sigma})).
\end{eqnarray} \end{widetext}Where, \bea
 {{{\cal{D}}}}&=&a_ka_{-k}+b_kc_{-k}+c_kb_{-k}=K_1^- +K_2^-
 +K_3^-
 \nn\\
 {{{\cal{D}}}}^{\dagger}&=&a^{\dag}_{-k}a^{\dag}_k+c^{\dag}_{-k}b^{\dag}_k+b^{\dag}_{-k}c^{\dag}_k=K_1^+ +K_2^+
 +K_3^+
 \nn\\
 N&=&\frac{1}{2}\{a^{\dag}_ka_k+a^{\dag}_{-k}a_{-k}+b^{\dag}_kb_k+b^{\dag}_{-k}b_{-k}+c^{\dag}_kc_k+c^{\dag}_{-k}c_{-k}+3\}=K_1^0+K_2^0+K_3^0 \nn \\
\eea  satisfy a $su(1,1)$ algebra
\begin{widetext}
\be
[N,{{{\cal{D}}}}]=-{{{\cal{D}}}};\;\;\;\;
[N,{{{\cal{D}}}}^{\dag}]={{{\cal{D}}}}^{\dag};\;\;\;
[{{{\cal{D}}}}^{\dag},{{{\cal{D}}}}]=-2N \ee. We also define:
 \bea
 {{\cal{D}}}_{\Sigma}&=&d_kd_{-k}
 \nn\\
 {{{\cal{D}}}}^{\dagger}_{\Sigma}&=&d^{\dag}_{-k}d^{\dag}_k
 \nn\\
 N_{\Sigma}&=&\frac{1}{2}\{d^{\dag}_kd_k+d^{\dag}_{-k}d_{-k}+1 \}
\eea \end{widetext} which also satisfy an su(1,1) algebra.

Thus, the Hamiltonian is linear in the generators of the Lie
Algebra of the $SU(1,1)$ group and following \cite{perel} the
solution of the time dependent Shrodinger equation
\begin{widetext}
\be
i\frac{d \psi(t)}{dt}=\int \momvol\frac{1}{a^3} 2\Omega_{\pi
}((\mu^2+\nu^{2})N+
\mu\nu({{\cal{D}}}+{{\cal{D}}}^{\dag}))\psi(0)\ee \end{widetext}
for each mode k
, we take $tanh(r_{k})$  for both
charged and neutral modes. 
Then we
have 
$|\psi(t)>=\Pi_{k}|\psi(k,t)>$ and with discrete k we have
for each mode k:
\be
i\frac{d \psi_k(t)}{dt}=H_{k} \psi_k(t)\ee,
where
\be
H_{k}=\frac{1}{a^3} 2\Omega_{\pi
}(k,t)((\mu^2+\nu^{2})N+
\mu\nu({{\cal{D}}}+{{\cal{D}}}^{\dag})).\ee
The solution $\psi_k(t)$
is given by the 
coherent state relevant to the discrete series
representation of the SU(1,1) group given by:
\be
\psi_k(t)=e^{i\phi t}|\alpha_k(t)>,\ee 
where,
\be
|\alpha_k(t)>= e^{\alpha_k {\cal {D_k}}^{\dag}}e^{\eta N_k} e^{
\alpha'(k) {\cal{D_k}}}|\psi(0)>.
 \ee
Here,$\alpha_k=tanh(r(k))$ and $\eta_k=2 ln (Cosh(r(k)))
=-ln(1-|\alpha_k|^2)$,$\alpha'(k)=-\alpha_k^{*}$.

$r(k)$ is
related to the frequencies $\Omega_{\pi}(k,t)$ and $\omega_{\pi}(k)$ by \be
Tanh(2r_k)= \frac{(\frac{\Omega_{\pi}(k,t)}{\omega_{\pi}})^2-1}{(\frac{\Omega_{\pi}(k,t)}{\omega_{\pi}})^2+1}\ee 

Where $\Omega_{\pi}(k,t\longrightarrow\infty)=\omega_{\pi}(k)$.
Thus in the evolution of the condensate , it is the  frequency
changes which  bring about squeezing \cite{Lo}.

From the definitions of $\cal{D}$,$\cal{D}^\dag$ and $N$, For
non-zero momenta, each of these exponentials has the form of a
two-mode squeezed state which, for two generic modes a and b, is
defined through [\cite{Barnett-Radmore}]:\begin{widetext}
\be
S_{ab}(\alpha)=e^{-a^{\dag} b^{\dag}
e^{(i\phi)}tanh(r)}e^{-(a^{\dag} a +b b^{\dag}) ln(cosh(r))}e^{b a
e^{(-i\phi)}tanh(r)}\ee \end{widetext} with the two-mode squeezed state being
given by: \be |\alpha_{ab}>=S_{ab}(\alpha)|0_{a};0_{b}>.\ee Using
the above we get

\begin{widetext}
\be
|\psi(k,t)>=(sech(r_{k}))^{3}\sum_{n,m,p=0}^{\infty}
\alpha_{k}^{n+m+p}
|n_{a_{k}},n_{a_{-k}}>|m_{b_{k}},m_{c_{-k}}>|p_{c_{k}},p_{b_{-k}}> \label{tm}
\ee
\end{widetext}

We denote by
 $n_0,n_+$ and $n_-$  the  number of produced $\pi_0's,\pi_+'s$ and $\pi_-'s$.
 At non-zero momentum, we have two such sets of number states, 
 one for forward momentum (k) and one for backward momentum (-k).

The states contributing to \ref{tm} are only the states which have
the same number of modes i.e $n_{a_{k}}=n_{a_{-k}}$,
$m_{b_{k}}=m_{c_{-k}}$ and $p_{c_{k}}=p_{b_{-k}}$. For a complete
statistical distribution of the pions  we find the p-ordered
characteristic function of the pion number distributions. this
will allow us to construct all the correlation functions. The
characteristic function of the neutral pion fields is
\begin{widetext}
\be
C_1(\xi,\eta,p)=<0_{a_k};0_{a_{-k}}|S_{a_k,a_{-k}}^{\dag}(r)e^{(\xi
a_{k}^{\dag}-\xi^*a_{k})} e^{(\eta
a_{-k}^\dag-\xi^*a_{-k})}S_{a_k,a_{-k}}(r)|0_{a_k};0_{a_{-k}}>e^{p(|\xi|^2+|\eta|^2/2)}
\ee 
\end{widetext}.

From this function it is relatively easy to construct all the
"p-th ordered moments" of the pion field through the formula
\begin{widetext}
\be
<a_{k}^{\dag q} a_{-k}^{\dag l}a_{-k}^{m}a_{k}^{n}>_{p}=
(\frac{\delta}{\delta\xi})^{q}
(\frac{\delta}{\delta\eta})^{l}(\frac{-\delta}{\delta\eta^{*}})^{m}(\frac{-\delta
 }{\delta\eta^{*}})^{n}C_1(\xi,\eta,p)|_{\xi=\eta=0} \ee 
 \end{widetext}.
Similarly for the charged
pions the Characteristic function is
\begin{widetext}
\be C_2
=(\xi,\eta,p)=<0_{b_k};0_{c_{-k}}|S_{b_k,c_{-k}}^{\dag}(r)e^{(\xi
b_{k}^\dag-\xi^*b_{k})} e^{(\eta b
c_{-k}^\dag-\xi*c_{-k})}S_{b_k,c_{-k}}|0_{b_k};0_{c_{-k}}>e^{p(|\xi|^2+|\eta|^2/2)}
\ee \end{widetext} and the correlation function is
\begin{widetext}
\be
<b_{k}^{\dag q} c_{-k}^{\dag l}c_{-k}^{m}b_{k}^{n}>_{p}=
(\frac{\delta}{\delta\xi})^{q}
(\frac{\delta}{\delta\eta})^{l}(\frac{-\delta}{\delta\eta^{*}})^{m}(\frac{-\delta
 }{\delta\eta^{*}})^{n}C_2(\xi,\eta,p)|_{\xi=\eta=0} \ee \end{widetext}
Thus for example the forward backward correlation functions for
neutral pions at non zero momenta are given by \be
<a_{k}a_{-k}>=sinh(r)cosh(r)=<a_{k}^{\dag}a_{-k}^{\dag}>\ee
Thus ,there is an entanglement of the forward and backward neutral
pions .
\be
<b_{k}c_{-k}>=sinh(r)cosh(r)=<b_{k}^{\dag}c_{-k}^{\dag}>\ee
\be
<c_{k}b_{-k}>=sinh(r)cosh(r)=<b_{-k}^{\dag}c_{k}^{\dag}>\ee
This entanglement provides us with correlations between the pions
with forward and backward momenta for both the charged and neutral
sector, the forward momentum positively charged pions with the
backward momentum negatively charged pions  and the backward
momentum positively charged pions and the forward momentum
negatively charged pions. 

In the non-zero k limit however the
distributions of the neutral and charged pions are the same. This
situation changes dramatically  in  the special case of soft pions
$k \longrightarrow 0$.
We will now show that in this limit the correlations of the
neutral pions as well as the number distributions are different in
the limit of large squeezing . To see this  observe that in  this
limit the probability distribution splits up into the convenient
form \begin{widetext}\be |<n_{0},n_{+},n_{-}|\psi(t)_{k=0}>|^{2}=|
<n_{0}|e^{r_0(a_{0}^{\dag})^{2}-r_0*a_{0}^{2}}|0>
<n_{+},n_{-}|e^{2r*_0 b ^{\dag}c^{\dag}-r*_0  bc}|0>|^{2} \ee \end{widetext}
where $r_0=lim_{k\longrightarrow 0} r(k)$ and $r_0^{*}$ is the
complex conjugate of $r_0$. Defining $S(r_0)$ as the one mode
squeezing operator

\begin{eqnarray}
S(r_0)&=&
<n_{0}|e^{r_0 (a_{0}^{\dag})^{2}-r_0^{*}a_{0}^{2})}|0>
\nonumber \\ &=& S_{n_{0},0}.
\end{eqnarray}
$S_{n_{+},n_{-},0}$ is then the two mode squeezing operator

\begin{eqnarray}
<n_{+},n_{-}|e^{(r_0 b^{\dag}c^{\dag}-r*_0^{*}bc)}|0> \nonumber \\
=S_{n_{+},n_{-},0}.
\end{eqnarray}

The neutral and charged pion distribution is:
\begin{equation}
P_{n_0,n_c}=<S_{n_{0},0}>^{2}<S^{\dag}{}^{m}_{n_{+},n_{-},0,0}>^{2}
\end{equation}
which is just the product of squeezed distributions for charged
and neutral pions and only even number of pions emerge. Writing
$n_{+}=n_{-}=n_{c}$, we get the distribution of charged particles
to be
\begin{equation}
P_{n_{c}}{}= \frac{(tanh(r_0))^{2n_{c}}}{(cosh(r_0))^{2}}
\end{equation}
\begin{widetext}
\begin{figure}[htbp]
\epsfxsize=7cm\epsfbox{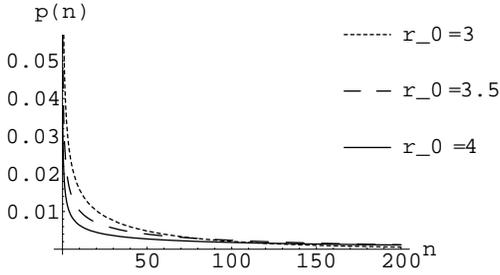} 
 \caption{Shows the variation of $P_{n_c}$ with $n_c$ for the $r_{0}=3,3.5$ and $ 4$}
\end{figure}
\end{widetext}

While the distribution of neutral pions is
\be
P_{n_{0}}=\frac{n_{0}!(tanh(r_0))^{n_{0}}}{((\frac{n_{0}}{2})!)^{2}cosh(r_0)
2^{n_{0}}} \ee
\begin{widetext}
\begin{figure}[htbp]
\epsfxsize=7cm\epsfbox{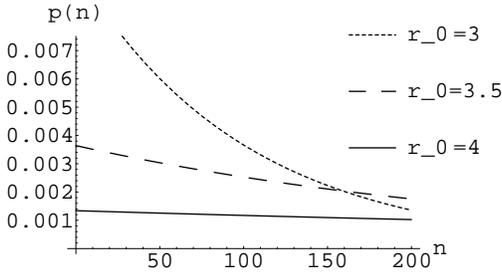} 
 \caption{Shows the variation of $P_{n_0}$ with $n_0$ for the $r_{0}=3,3.5$ and $ 4$}
\end{figure}
\end{widetext}

Comparing the charged and neutral pion distributions for a
squeezing parameter  which corresponds to the total no. of pions
$<n>=40$ ,(Fig. 3) and a squeezing parameter which
corresponds to $<n>=2400$ (Fig.4)we see that there is a marked
difference in the charged and neutral pion distributions for large
values of the squeezing parameter for pions at zero relative
momenta(soft pions).
\begin{widetext}
\begin{figure}[htbp]
\epsfxsize=7cm\epsfbox{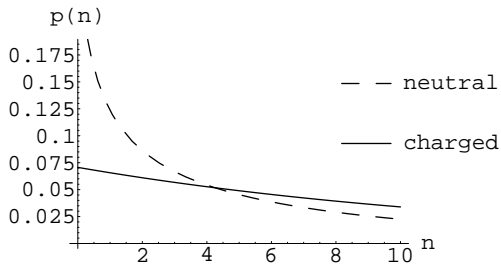} 
 \caption{Shows the comparison of  $P_{n_0}(n)$ and $P_{n_c}(n)$ for the $<n>=40$ }
\end{figure}
\end{widetext}
\begin{figure}[htbp]
\epsfxsize=7cm\epsfbox{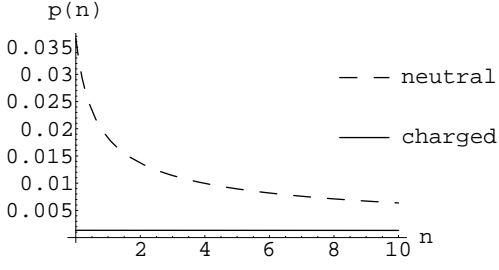}  
\caption{Shows the comparison of  $P_{n_0}(n)$ and $P_{n_c}(n)$ for the $<n>=2400$ }
\end{figure}
In  I  we  have found in the evolution of
this Hamiltonian  that in the quenched limit ( fast expansion) the
low momentum modes are enhanced significantly, whereas in the
adiabatic (slow expansion) limit, no such enhancement occurs. This
has been shown to be directly related to the value of the
squeezing parameter, since for each mode $<n>=Sinh^2(r_0)$.
 The difference in the total charged and multiplicity
 distributions of pions produced at low momenta (soft pions)
for large squeezing parameter is a direct mirror of this effect.
Thus the measurable quantities like the charged and neutral total
multiplicity distributions at low momenta can be used to determine
whether a quench or and adiabatic expansion occurred in a heavy
ion collision and whether a disoriented chiral condensate formed
or not. This serves as a very useful signal for non-equilibrium
phase transitions in Heavy ion collisions.
We now proceed to examine whether this dramatic effect survives
theoretically the imposition of isospin conservation.
\newpage
\section{The Isospin Symmetry of the Hamiltonian}
We now examine the isospin structure of the pion Hamiltonian.
Since isospin is conserved in strong interactions we need to
extract states of fixed isospin from the squeezed number states
which will be the eigenstates of the su(1,1) Hamiltonian
constructed above. Thus far, most studies have constructed such
isospin states at zero momentum. However, for an event by event
analysis  we need to  construct these states at non-zero momentum
states which have not been considered before. We will do so by
 defining the  isospin operators
constructed from the Fock state basis of pions at non zero
momentum by constructing the vectors
 \bea {\vec{\it{a}}}(k)=
 \lrb \ba{c} \frac{b_k+c_k}{\sqrt{2}}\\ \frac{i(b_k-c_k)}{\sqrt{2}}\\a_k \ea \rrb =
\lrb \ba{c}a_1(k)
\\a_2(k) \\ a_3(k) \ea \rrb
\eea  \bea {\vec{\it{a}}}(k)=\lrb \ba{c}
\frac{b_{-k}+c_{-k}}{\sqrt{2}}\\
\frac{i(b_{-k}-c_{-k})}{\sqrt{2}}\\a_{-k} \ea \rrb = \lrb
\ba{c}a_{1}(-k)
\\ a_{2}(-k)\\ a_{3}(-k) \ea \rrb
\eea From the above ,two sets of isospin operators can be
constructed

\bea
 I_{i}(k)&=&i\epsilon_{ijl}a_j(k)a_l(k)^{\dag} \\ \nn
 I_{i}(-k)&=&i\epsilon_{ijl}a_j(-k)a_l(-k)^{\dag}.
\eea

Generalizing the work of \cite{skagerstam} , we find that the two operators
$I_i(k)$and $I_i(-k)$ taken together  define a direct sum algebra
of two su(2) algebras su(2)+su(2)=so(4)defined by algebra
 the generators \bea M_i&=&I_i(k)+I_i(-k)
\nn \\ N_i&=&I_i(k)-I_i(-k), \eea which satisfy the following relations
\begin{widetext}
\be
[M_i,N_j]=i\epsilon_{ijl}M_l;\;\;\;\;
[N_i,N_j]=i\epsilon_{ijl}M_l;\;\;\; [M_i,M_j]=i\epsilon_{ijl}M_l
\ee 
\end{widetext}characteristic of  an so(4) algebra. The Casimir operators of
this algebra are \bea
        C_1=  M^2+N^2&=&I(k)^2+I(-k)^2 \\ \nn
         C_2=  M\bullet N&=&I(k)^2-I(-k)^2
           \eea

From the vectors $\vec{\it{a}}(k)$ and $\vec{\it{a}(-k)}$ we can
form bilinear operators
 \bea{\cal{A}}(k)&=&\vec{\it{a}(k)}\cdot\vec{\it{a}(k)} \nn \\
{\cal{A^{\dag}}}(k)&=&\vec{\it{a}^{\dag}(k)}\cdot\vec{\it{a}^{\dag}(k)} \nn \\
{\it{N(k)}}&=&\vec{\it{a}^{\dag}(k)}\cdot\vec{\it{a}(k)} \eea
which satisfy the mdified su(1,1) algebra:
 \bea
\lsb{\it{N(k)}},{\cal{A}}(k)\rsb&=&-{{{\cal{A}}}}(k);\;\;\;\;\nn \\
\lsb{\it{N(k)}},{\cal{A}}(k)^{\dag}\rsb&=&{\cal{A}}(k)^{\dag};\;\;\;\nn
\\ \lsb{\cal{A}}^{\dag}(k),{\cal{A}}(k)\rsb=-4{\it{N(k)}}-6 \nn\\
\eea

It is pertinent to note here that the relationship between the $su(1,1)$ algebra defined earlier and the so(4) algebra is \begin{widetext}
 \be D={\vec{\it{a}}}(k){\vec{\it{a}}}(-k) \;\;\;
 D^{\dag}={\vec{\it{a}}}^{\dag}(-k){\vec{\it{a}}}^{\dag}(k)\;\;N=\frac{1}{2}(\vec{\it{a}}^{\dag}(k)\vec{\it{a}}(k)+{\vec{\it{a}}}^{\dag}(-k){\vec{\it{a}}}(-k)+3) \ee
\end{widetext}
Since the two algebras are related , we can also relate the
eigenstates of H constructed through these two algebras.

Since $[H,M_3]=0$ $[H,N_3]=0$ and $[H,C_i]_{i=1,2}=0$,  the eigenstates of
$M_3, N_3$ and $C_i$ are
 defined by a generalization of the spherical harmonics
  $
 Y^{l}_{m}({\it{a}}(k)^{\dagger})
Y^{l}_{m}({\it{a}_(-k)}^{\dagger})|0>$ which are  also eigenstates
 of H.

In terms of the Isospin operators the pion Hamiltonian can be
written as:\begin{widetext}
\begin{equation}
H_{\pi}=\int \momvol\frac{1}{a^3} 2\Omega_{\pi
k}(\mu^2+\nu^{2})\frac{1}{2}(\it{a}^{\dag}(k)\it{a}(k)+{\vec{\it{a}}}^{\dag}(-k){\vec{\it{a}}}(-k)+3)
+2\Omega_{\pi
k}\mu\nu({\vec{\it{a}}}(k){\vec{\it{a}}}(-k)+{\vec{\it{a}}}^{\dag}(-k){\vec{\it{a}}}^{\dag}(k))
\end{equation}
\end{widetext}
Since the pion Hamiltonian can be written both in terms of su(1,1)
and isospin generators the eigenstates of the Hamiltonian have a
Janus faced nature and can be both su(1,1) eigenstates and
su(2)xsu(2) eigenstates. This allows the calculation of
multiplicity distributions of pions in the number state basis as
well as in the isospin basis as we shall see in the next section.
\section{Isospin Squeezed states for $k\longrightarrow 0$ pions}
First, we construct the isospin squeezed states for the
$k\longrightarrow 0$ case. For this we need to relate the states
\begin{widetext}
\be
|r_0>=e^{r_0({\cal {D}}^{\dag}-{\cal {D}})}|\psi_0(0)>=
e^{r_0(a_0^{\dag^2}+c^{\dag}_{0}b^{\dag}_{0}+b^{\dag}_{0}c^{\dag}_{0}-a_0^2+b_{0}c_{0}+c_{0}b_{0})}|\psi_0(0)>
\ee 
\end{widetext}
to the states of definite isospin. Note that we have
added a subscript $0$ to the operators to indicate the zero
momentum limit.

For this purpose we note that for $k-->0$ there exists only one
isospin algebra constructed from the operator \bea
{\vec{\it{a}}}_0=
 \lrb \ba{c} \frac{b_0+c_0}{\sqrt{2}}\\ \frac{i(b_0-c_0)}{\sqrt{2}}\\a_0 \ea \rrb
 \eea
This algebra is given in terms of \be
 I_{i0}=i\epsilon_{ijl}a_{j0}a_{l0}^{\dag}
 \ee
 The squared Isospin vector is
 \be
 I^2=N^2+N+{\cal{A}}_{0}^{\dag}{\cal{A}}_{0}
\ee For further discussion, now, in the $k-->0$ case we drop the
subscript $0$ . The definitions are \bea
{\cal{A}}^{\dag}&=&\vec{\it{a}}^{\dag}\bullet
\vec{\it{a}}^{\dag}\\ \nn {\cal{A}}&=&\vec{\it{a}}\bullet
\vec{\it{a}} \eea We also have the relations \bea \lsb
{\cal{A}},Y_l^m(\vec{{\it{a}}^{\dag}}) \rsb&=&0 \\ \nn \lsb
N,{\cal{A}} \rsb&=&-2{\cal{A}}
\\ \nn \lsb N,A^{\dag}\rsb&=&2{\cal{A}}^{\dag} \eea \begin{widetext}
\bea Y^{l}_{m}(\vec{a}^{\dag})|0>&=&(- 2)^{-m}\sqrt{(2l+1) (l-m)!
(l+m)!} \\ \nn &\sum_{n=0}^{(l-m)/2}&
\frac{2^{m/2-n}}{(l-m-2n)!n!(n+m)!}(a_0)^{l-m-2n}(a_{+})^{n+m}(a_{-})^{n}|0>\eea
\end{widetext}
are the eigenstates of $I^2$, $I_3$ and $N$ whose action on them
is given by: \bea
|l,m,l+2n>&=&N_{l,n}({\cal{A}}^{\dag})^nY^{l}_{m}(\vec{{\it{a}}^{\dag}})|0>
\\ \nn
I^2|l,m,l+2n>&=&l(l+1)|l,m,l+2n> \\ \nn
I_3|l,m,l+2n>&=&m|l,m,l+2n> \\ \nn N|l,m,l+2n>&=&(l+2n)|l,m,l+2n>
\eea and the relationship between the number states $|n_0,n_+,
n_->$ and $|l,m,l+2n> $ is given by
\begin{widetext}
\be
|l,m,l+2n>=\sum_{p=0}^{n+(l-|m|)/2}c^{(l,m,n)}(p)|l+2n-|m|-2p,\frac{m+|m|}{2}+p,\frac{m-|m|}{2}+p>
\ee
\end{widetext}
\begin{widetext}
 \bea c_{p}^{l,m,n}=\left( -1 \right) ^m\,
  {\sqrt{\frac{2^{\frac{l - m}{2} + p}\,
        \left( l + n \right) \,\left( 1 + 2\,l \right) !\,
        \left( l + m \right) !\,n!\,
        \left( l - m + 2\,n - 2\,p \right) !\,
        \left( m + p \right) !}{\left( l - m \right) !\,
        {m!}^2\,\left( 1 + 2\,l + 2\,n \right) !\,
        {\left( n - p \right) !}^2\,p!}}} \nn\\\sum_{j = 0}^{p}
    \frac{\frac{{\left( -1 \right) }^j}{2^{2\,j}}}
     {\left( p - j \right) !\,
       \left( n + j - p \right) !\,
       \left( l - 2\,j \right) !\,j!\,j!} \nn \\
        \eea

\end{widetext}

 The inverse relation is
\begin{widetext}
\be
|n_0,n_+,n_->=\sum_{l=0}^{n_++n_-+n_0}c^{(l,n_+-n_-,n_++n_-+n_0)}_{n_++n_--\frac{|n_+-n_-|}{2}}|l,n_+-n_-,n_++n_-+n_0>
\ee
\end{widetext}

For strong interactions involving  charged pions $n_+-n_-=0$ by
charge conservation , hence the only values of $c_{p}^{l,m,n}$
contributing to physically measurable quantities will be
\begin{widetext}
 \bea
c_{(n_++n_-)}^{l,0,n}&=&2^{(n_++n_-+\frac{1}{2}l)}
(\frac{(n+l)!(2l+1)!l!l!(l+2n-2(n_++n_-))!(n_++n_-)!n!(n_++n_-)!}{(2n+2l+1)!})^{\frac{1}{2}}
 \nn  \\&\sum_{j=0}^{(n_++n_-)}& \frac{{\left( -1 \right) }^j}
  {(2^{2\,j})\,j!\,\left( -j + (n_++n_-) \right) !\,
    \left( -2\,j + l \right) !\,
    \left( j \right) !\,\left( j - (n_++n_-) + n \right) !}
\eea
\end{widetext}

In considering the probability distribution of pions with isospin
conservation the squeezed state of definite isospin has to be
projected from the number state distribution . \be<n_0,n_+,n_-|S|l,m,2n>=\sum_{p=0}^{n+(l-|m|)/2}c^{(l,m,n)}_{p}
<n_0,n_+,n_-|S|l+2n-m-2p,m+p,+p>
\ee 

Where we have assume that $m\geq0$.

 The general expression
for the one mode and two mode matrix elements are \cite{mvs}:
\begin{widetext}
\begin{eqnarray}
S_{n_0,l+2n-m-2p}^{one
mode(r)}&=&(-1)^{\frac{l+2n-m-2p+n_0}{2}}(\frac{l+2n-m-2p!
n_0!}{cosh(r)})(\frac{tanh(r)}{2})^{\frac{l+2n-m-2p+n_0}{2}}\nonumber
\\
&&\sum_{\lambda}^{min[\frac{n_0}{2},\frac{l+2n-m-2p}{2}]}\frac{(\frac{-4}{sinh^{2}\eta})^{\lambda}}{(2\lambda)!
(\frac{l+2n-m-2p}{2}-\lambda)!(\frac{n_0}{2}-\lambda)!}
\end{eqnarray}
\end{widetext}
$n$,$m$, even.\begin{widetext}\begin{eqnarray}
S^{Two-mode}_{n_{+},n_{-},m+p,p}&=&(-1)^{2n_{+}+m+p}\sqrt{n_{+}!n_{-}!(m+p)!(p)!}\frac{(tanh(r))^{n_{+}+m+p}}{(cosh(r))^{n_{-}-n_{+}+1}}\nonumber
\\ &&\times
\sum_{\lambda}\frac{(\frac{1}{sinh^{2}(r)})^{\lambda}}{\lambda!(n_{+}-\lambda)!(m+p-\lambda)!(n_{-}-n_{+}+\lambda)!}
\end{eqnarray}
\end{widetext}

Using these identities we have
 \begin{widetext}
\be
<n_0,n_-,n_+|S(r)|l,m,2n>=
\sum_{p=0}^{n+(l-|m|)/2}c^{(l,m,n)}_{p}S_{n_0,l+2n-m-2p}^{onemode}(r)S^{Two-mode}_{n_+,n_-,m+p,p}(r) \ee 
\end{widetext}

 For the
special case of pion emission in strong interactions ,
 the charge
is conserved, hence $ m=n_+-n_-=0$ .

To show clearly the effect of isospin conservation , we focus on total isospin $I=0$ as an example.
In this case,
\begin{widetext}
\be
<0|S|n_0,n_c>=\frac{(-tanh(r))^{n_0+n_c}}{(cosh(r))^{3/2}} 
(\frac{n_0+n_c!}{(2(n_0+n_c)
+1)!)})^{\frac{1}{2}}2^{n_c} (\frac{(2n_0)!}{(n_0)!}), \ee \end{widetext}with
$n_++n_-=n_c$ and $n=n_0+n_c$.

Thus, the probability of finding $n_0$ neutral pions and $n_c$
charged pions in the state of I= 0 is

\be
P_{n_0,n_c}=\frac{(tanh(r))^{(n_0+n_c)}}{(cosh(r))^{3}}(\frac{((n_0+n_c))!}{(2(n_0+n_c)+1)!)}4^{n_c}
\frac{(2n_0)!}{(n_0! )^2}\ee This gives the neutral pion
distribution $P_{n_0}=\sum_{n_c}P_{n_0,n_c}$

 as \begin{widetext}\be P_{n_0}= \frac{2^{-1 - 2\,n}\,e^{{\tanh
(r)}^2}\,{\sqrt{\pi }}\,{{Csch}(r)}^2\,\left( 2\,n \right) !\,
    \left( {\Gamma}(\frac{1}{2} + n) - {\Gamma}(\frac{1}{2} + n,{\tanh (r)}^2) \right) \,
    {\left( {\sinh (r)}^2 \right) }^n\,{\left( {\tanh (r)}^2 \right) }^{\frac{1}{2} - n}}{{\left( {\cosh (r)}^2 \right) }^n\,
    {n!}^2\,{\Gamma}(\frac{1}{2} + n)}\ee \end{widetext}
    In terms of the average number $<n_0>$ of neutral pions
\begin{widetext} \be P_{n_{0}}= \frac{2^{-1 - 2\,n_0}e^
     {\frac{<n_0>}{1 + <n_0>}}\,
    {<n_0>}^{-1 + n_0}\,
    {\left( \frac{<n_0>}{1 + <n_0>}
        \right) }^{\frac{1}{2} - n_0}\,{\sqrt{\pi }}\,
    \left( 2\,n_0 \right) !\,
    \left( \Gamma(\frac{1}{2} + n_0) -
      \Gamma(\frac{1}{2} + n_0,
       \frac{<n_0>}{1 + <n_0>}) \right) }
    {{\left( 1 + <n_0> \right) }^{n_0}\,{n_0!}^2\,
    \Gamma(\frac{1}{2} + n_0)} ,\ee \end{widetext}

     The  coresponding distribution of charged pions is given by
\begin{widetext}\be P_{n_c}=\frac{{\sqrt{\pi
}}\,1F_1(\frac{1}{2},\frac{3}{2} + n,{\tanh (r)}^2)\,{Sech(r)}^2\,
    {\left( {\sinh (r)}^2 \right) }^n}{2\,{\left( {\cosh (r)}^2 \right) }^n\,\Gamma(\frac{3}{2} +
    n)}\ee \end{widetext}
In terms of $<n_c>$ we have \begin{widetext}
 \be P_{n_c}(<n>)= \frac{{<n>}^{n_c}
    {\left( 1 + <n> \right) }^{-1 - n_c}
    {\sqrt{\pi }}^1F_1(
     \frac{1}{2},\frac{3}{2} + n_c,
     \frac{<n>}{1 + <n>})}{2
    \Gamma(\frac{3}{2} + n_c)})\ee \end{widetext}
\begin{widetext}
\begin{figure}[htbp]
\epsfxsize=7cm\epsfbox{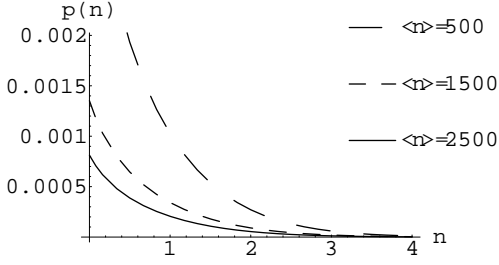}  \caption{Shows the
variation P(n) with n with isospin $I=0$ for NEUTRAL PIONS for
values of $<n>=500, 1500,2500.$}
\end{figure}
\end{widetext}
\begin{widetext}
\begin{figure}[htbp]
\epsfxsize=7cm\epsfbox{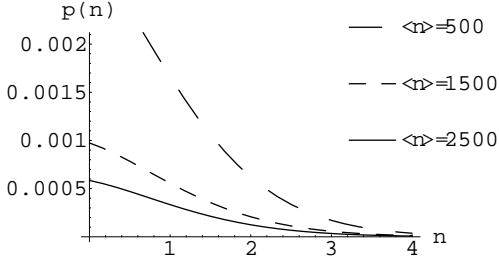}  
\caption{Shows the variation P(n) with n for CHARGED PIONS for values of $<n>=500$,
$1500$,$2500$.}\end{figure} \end{widetext}

To compare neutral and charged pion distributions for large and
small squeezing parameters the following figures are presented.
Fig.[6] shows the variation of $P_{n_c}$ with $n_c$ for various values of $<n>$.
\begin{figure}
\epsfxsize=7cm\epsfbox{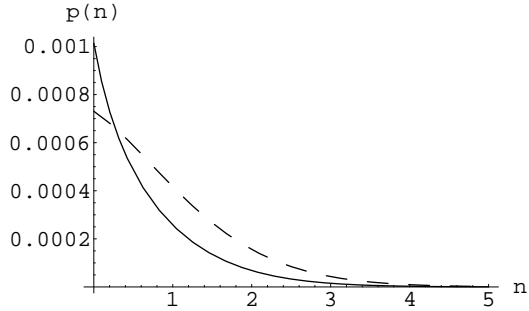}  
\caption{Shows the variation P(n) with n for CHARGED PIONS (dashed line) and neutral
pions (solid line) for values of $r_0=3.5$}
\end{figure}
\begin{figure}
\epsfxsize=7cm\epsfbox{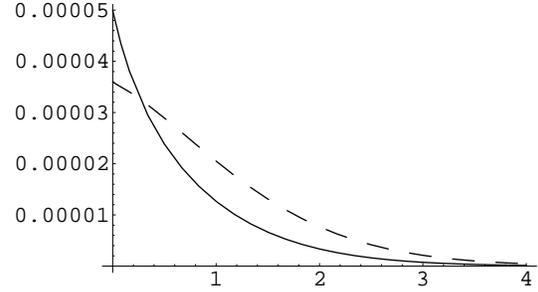}  
\caption{Shows the variation P(n) with n for CHARGED PIONS (dashed line) and
neutral pions (solid line) for values of $r_0=5$}
\end{figure}
Comparing the plots of the difference between charged and neutral
pions distributions without isospin (fig.4) and those with
isospin conservation (fig.7), we find that the dramatic
difference in the two distributions is reduced by the imposition
of isospin conservation for $I=0$. However, still the charged pion
distributions get broader with increasing squeezing and hence the
distributions can still be used to determine whether a phase
transition has occurred in Heavy ion collisions even when isospin
conservation for soft pions is taken into account.

\newpage

\section{Isospin squeezed states for pions with arbitrary momentum}
In this section we calculate the multiplicity
distributions of squeezed pion states at non-zero momenta with the
imposition of Isospin invariance. For arbitrary k we have to
proceed slightly differently from above. We have seen that the
states 
\be |\alpha>= N e^{\alpha D^{\dag}}|0> \ee 
represent the dynamical wavefunctions of the pions at non zero momenta where,
$D^{\dag}(k)=\alpha{\vec{\it{a}}}^{\dag}(-k){\vec{\it{a}}}^{\dag}(k)$.  
Using the identitity:
\begin{widetext}
\be
e^{k\cdot x}
=\sum_{l=0}^{\infty}\phi_{l}(k^2x^2)\sum_{-l}^{+l}Y_{l}^{m}(k)^{*}Y_{m}^{l}(x),
\ee
\end{widetext}
where $\phi_l$ is the spherical Bessel Function $\phi_l(x)=\frac{j_{l}(-i\sqrt{x})}{(\sqrt{x})^l}$,
and
\be
\phi_l(x)=2^l\sum_{n=0}^{\infty}\frac{(n+l)!x^n}{(2n+2l+1)!n!}. \ee

We write
\begin{widetext}
\be
|\alpha>=Ne^{\alpha{\vec{\it{a}}}^{\dag}(-k){\vec{\it{a}}}^{\dag}(k)}|0>_{k}|0>_{-k}=N
\sum_{l=0}^{\infty}\phi_{l}(\alpha^2{\cal{A}}(k)^{\dag}{\cal{A}}(-k)^{\dag})\sum_{m=-l}^{+l}Y_{l}^{m}(\sqrt{\alpha}\vec{\it{a}(k)}^{\dag})^{*}Y_{m}^{l}(\sqrt{\alpha}\it{a}(k)^{\dag})|0>_k|0>_{-k}
\ee 
\end{widetext}

To get the isospin decomposition we use the fact that
\begin{widetext}
 \bea Y^{l}_{m}(\vec{{\it{a}}(k)^{\dag}})|0,0,0>&=&\\ \nn
 &(- 2)^{-m}&\sqrt{(2l+1) (l-m)!
(l+m)!}\\ \nn
&\sum_{n=0}^{(l-m)/2}&\frac{2^{m/2-n}}{(l-m-2n)!n!(n+m)!}(a_{k,0})^{l-m-2n}(a_{k+})^{n+m}(a_{k-})^{n}|0,0,0>.\eea
\end{widetext}

 $Y_l^{m}(\vec{{\it{a}}(k)^{\dag}})|0>$ is an
eigenstate of $I_k^2$ and $I_{3k}$ and correspondingly
$Y_l^{m}(\vec{{\it{a}}(-k)^{\dag}})|0>$ is an eigenstate of
$I(-k)^{2}$ and $I_{3-k}$.

The number operator eigenstates and the isospin eigenstates are related by the formula:
\be|l,m,l+2n(k)>=N_{l,n}({\cal{A}}(k)^{\dag})^{n(k)}Y_l^m(\vec{a_k}^{\dag})|0>_{k}
\ee
where $N_{l,n}=(\frac{2^l(n+l)!}{(2n+2l+1)!n!)})^{\frac{1}{2}}$

Thus, the squeezed state wave function for fixed isospin is given by
\begin{widetext}
\be e^{\alpha D^{\dag}}|0>_{k}|0>_{-k}=4 \pi
\sum_{l=0}^{\infty}\sum_{n=0}^{\infty}\sum_{m=-l}^{l}\alpha^{l+2n}\frac{({\cal{A}}(k)^{\dag})^{n}({\cal{A}}(-k)^{\dag})^{n}(n+l)!}{n!
(1+2n+2l)!}Y^{l*}_{m}(\vec{{\it{a}}(k)^{\dag}})Y^{l}_{m}(\vec{\it{a}}(-k)^{\dag})|0>_{k}|0>_{-k}
.\ee  \end{widetext}

Because, $Y^{l}_{-m}=(-1)^mY^l_m(\vec{e})^*$
and $Y_m^l(c\vec{e})=c^lY^l_m(\vec{e})$ we get \begin{widetext}
\be
|\alpha>=N_{l.n}\sum_{n=0}^{\infty}\sum_{l=0}^{\infty}\sum_{m=-l}^{l}\frac{\alpha^{2n+l}}{2^l(n+l)!}(-1)^{m}|l,-m,l+2n>_{(k)}|l,m,l+2n>_{(-k)}
.\ee  \end{widetext}
We now have the necessary ammunition to  calculate the multiplicity
distribution:
\be
P_{l,m,n}^{n_0,n_-,n_+}=<n_o,n_+,n_-|_k<n_0,n_-,n_+|_{-k}|\alpha>^2
.\ee 

Defining \be <n_o,n_+,n_-|_k|l,-m,l+2n>_{(k)}=
c_{p}^{l,m,n}(k)\ee and \be
<n_o,n_+,n_-|_{-k}|l,-m,l+2n>_{(-k)}= c_{p}^{l,m,n}(-k),\ee
and using the results of section III,
we write the total probability of finding a state of $2n=n(k)+n(-k)$
pions of momentum k and -k  of total Isospin I= l and $I_3=m$ as
\begin{widetext}
\be
P_{l,m,n}^{n_{0,\pm k} ,n_{-,\pm k} n_{+,\pm k}}
(\alpha)=(\frac{(tanh(r))^{n(k)+n(-k)+l}}{Sech(r)^{3/2}
2^l(n(k)+n(-k)+l)!}c_{p}^{l,m,n}(k)c_{p}^{l,m,n}(-k))^2 \ee
\end{widetext} 

For
I=0 and m=0 we have

\begin{widetext}
\be
 P_{0,0,n}^{n_{0}(\pm k) ,n_{-}(\pm k) n_{+}(\pm k) }(\alpha)
=(\frac{(tanh(r))^{2(n(k)+n(-k)+l)}}{Sech(r)^{3}
(n(k)+n(-k)+l)!}
(\frac{((n_0(k)+n_0(-k)+n_c(k)+n_c(-k)))!2^{n_0(k)+n_0(-k)+n_c(k)+n_c(-k)}}{(n_0(k)+n_0(-k)+n_c(k)+n_c(-k))+1)!})
\ee
\end{widetext}

In figure 9 we show the effect of isospin conservation at non
zero momentum on the total multiplicity distribution of pions and
see that it has a significant effect of narrowing the multiplicity
distribution. However as in the case without isospin conservation
at non zero momentum there is no difference between the neutral
and squeezed pion distributions.

\begin{widetext}

 \begin{figure}[htbp]
\epsfxsize=7cm\epsfbox{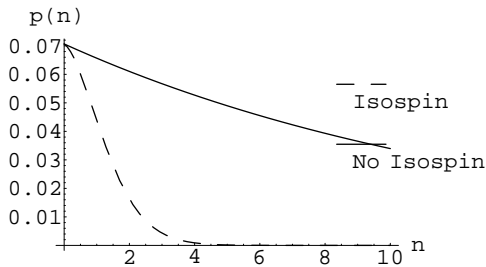}  \caption{Shows the
variation P(n) with n for pions with isospin conservation
(I=0)(dashed line) and without isospin conservation ( solid line)
for values of $r_0=5$}
\end{figure}
\end{widetext}

\section{Conclusion}
To conclude we have given a 
dynamical model for the production of pions 
in heavy ion collisions  going though a 
non-equilibrium phase transition, 
and have shown how the charged and neutral pion distributions
 both globally and on an event by event basis
 can be used as a signal to measure the nature of the phase transitions.
We have shown that the distributions of the charged and 
neutral zero momentum pions are conspicuously different 
when the system undergoes a quench
 as is thought to occur when a disoriented chiral condensate forms in a heavy ion collision.
We have shown that this dramatic difference disappears when non
 zero momentum number distributions of pions are observed.
However forward backard correlations are enchanced both in the
 neutral and charged pion sectors when the system undergoes a quench , which is parametrised by a squeezing parameter.
We have shown how to incorporate isospin conservation on both the zero momentum and the non-zero momentum pion distributions and shown that the effect of the quench survives the imposition of this invariance on the pion multiplicity distributions.
Thus we have presented a field theoretical dynamical description 
of the signals that can differentiate between the various ways in which a system can undergo a phase transition in heavy ion collisions.
 The distributions and correlations which we have derived in this paper should be useful in 
testing the data envisaged at the RHIC collider to discern 
not only whether a disoriented chiral condensate
 is formed by easily measurable pion distributions,
 but also the effect of conservation laws on the general distributions of pions emerging
in the collision processs\cite{tapan}.
It is our  hope that this work will prove useful to experimentalists when examining the data and we look forward to its application.


\begin{thebibliography}{141}
\bibitem{knox}
W.~J.~Knox,
``Effect Of Bose-Einstein Statistics On Multiplicity Distributions And Correlations In Multiparticle Production Processes At High Energies,''
Phys. Rev.D {\bf 10}, 65 (1974).
\bibitem{giov}A.~Giovannini and L.~Van Hove, Z.\ Phys.\ C {\bf 30}, 391 (1986).
\bibitem{carruthers}
P.~Carruthers and C.~C.~Shih,
``The Phenomenological Analysis Of Hadronic Multiplicity Distributions,''
Int.\ J.\ Mod.\ Phys.\ A {\bf 2}, 1447 (1987).
\bibitem{bambah} B.Bambah and M.V.Satyanarayana Phys.\ Rev.\ D {\bf 38}(1988) 2202.
\bibitem{weiner}
G.~N.~Fowler and R.~M.~Weiner, ``Application Of The Methods Of
Quantum Optics To Multi - Hadron Production,'' Adv.\ Ser.\
Direct.\ High Energy Phys.\  {\bf 2}, 481 (1988).
\bibitem{horn}D.Horn and R.Silver  Ann.\ Phys.  {\bf 66} (1971) 509.
\bibitem{botke}
J.~C.~Botke, D.~J.~Scalapino and R.~L.~Sugar,
``Coherent States And Particle Production,''
Phys.\ Rev.\ D {\bf 9} (1974) 813.
\bibitem{Waka}M.~Wakamatsu, Nuovo Cim.\ A {\bf 56}, 336 (1980).
\bibitem{biya}M.~Biyajima, T.~Kawabe and N.~Suzuki,
``Forward - Backward Multiplicity Distributions In The Perina-Mcgill Scheme And Analyses In E+ E- Collisions,''
Z.\ Phys.\ C {\bf 35}, 215 (1987).
\bibitem{Kogan:1993as}
I.~I.~Kogan,
JETP Lett.\  {\bf 59}, 307 (1994)
[arXiv:hep-ph/9310245].
\bibitem{kowal} K.L Kowalski and C.C. Taylor. A white paper for full acceptance detector , CWRUTH-92-6.
\bibitem{bj1} J.D.Bjorken Int.
Journal of Mod. Phys. {\bf A7}(1992)4189.
\bibitem{rw}
K.~Rajagopal and F.~Wilczek,
 ``Emergence of coherent long wavelength oscillations after a quench:
Application to QCD,''
Nucl.\ Phys.\ B {\bf 404}, 577 (1993)
[arXiv:hep-ph/9303281].
\bibitem{Mart}
M.~Martinis and V.~Mikuta, Phys.\ Rev.\ D {\bf 12}, 909 (1975).,
\bibitem{mart}M.~Martinis and V.~Mikuta-Martinis,
 '`Disoriented chiral condensate and charge-neutral particle fluctuations in
heavy ion collisions,''
arXiv:nucl-th/0309078.
\bibitem{Amado}
R.~D.~Amado and I.~I.~Kogan,
Phys.\ Rev.\ D {\bf 51}, 190 (1995)
[arXiv:hep-ph/9407252].
\bibitem{andreev} I.V. Andreev,Phys.Atom.Nucl. 63 (2000) 1988-1992; Yad.Fiz. 63 (2000) 2080-2084.
\bibitem{botke2}J.~C.~Botke, D.~J.~Scalapino and R.~L.~Sugar,
Phys.\ Rev.\ D {\bf 10} (1974) 1604.
\bibitem{skagerstam}K.Eriksson and B. Skagerstam J. Phys. A.  {\bf 12}(1979)2175.
\bibitem{hwa}I.~M.~Dremin and R.~C.~Hwa,
Phys.\ Rev.\ D {\bf 53} (1996) 1216 [arXiv:hep-ph/9510223].
\bibitem{Hiro} H.~Hiro-Oka and H.~Minakata,
Phys.\ Lett.\ B {\bf 425} (1998) 129 [Erratum-ibid.\ B {\bf 434}
(1998) 461] [arXiv:hep-ph/9712476].
\\
H.~Hiro-Oka,
{\it Prepared for TMU - Yale Symposium on Dynamics of Gauge Fields: An External Activity of APCTP, Tokyo, Japan, 13-15 Dec 1999}
\bibitem{suzuki}
M.~Suzuki,
Phys.\ Rev.\ D {\bf 52} (1995) 2982.
\bibitem{bm1}
B.~Bambah and C.~Mukku,
arXiv:hep-ph/0303086.
\bibitem{maedan}
S.~Maedan,
Phys.\ Rev.\ D {\bf 67}, 014003 (2003)
[arXiv:hep-ph/0209356].
\bibitem{Barnett-Radmore}S.M. Barnett, P.M. Radmore, "Methods in
Theoretical Quantum Optics" Clarendon Press, Oxford,1997.
\bibitem{mvs} M.V. Satyanarayana, Phys. Rev. D 32, 400–404 (1985)
\bibitem{perel}A.M. Perelomov , "GENERALIZED COHERENT STATES AND THEIR APPLICATIONS",
By (Moscow, ITEP),. 1986. 320pp.
Berlin, Germany: Springer (1986) 320 p.
\bibitem{bam96}B. Bambah in "Proceedings of the IV Int. Conf. on Squeezed States, Shanxi, China" NASA PUB. 3232,p3. [hep-ph/9604371]
\bibitem{Lo} C.F. Lo, Phys.Rev. A43, 404  (1991)
\bibitem{sigal}E.A. Kuraev, Z.K. Silagadze, Acta Phys.Polon. B34 (2003) 4019-4072.
\bibitem{tapan}B. Mohanty, T.K. Nayak, D.P. Mahapatra, Y.P. Viyogi
  nucl-ex/0211007.
\end{thebibliography}
\end{document}